\documentclass{vgtc}                          

\ifpdf
  \pdfoutput=1\relax                   
  \usepackage{graphicx}                
  \DeclareGraphicsExtensions{.pdf,.png,.jpg,.jpeg} 
\else
  \usepackage{graphicx}                
  \DeclareGraphicsExtensions{.eps}     
\fi%

\graphicspath{{figures/}{pictures/}{images/}{./}} 

\usepackage{microtype}                 
\PassOptionsToPackage{warn}{textcomp}  
\usepackage{textcomp}                  
\usepackage{mathptmx}                  
\usepackage{times}                     
\usepackage{cite}                      
\usepackage{tabu}                      
\usepackage{booktabs}                  
\usepackage{float}
\usepackage{amsmath}

\onlineid{0}

\vgtccategory{Research}

\vgtcinsertpkg

\title{Trust Calibration as a Function of the Evolution of Uncertainty in Knowledge Generation: A Survey}

\author{Joshua Boley\thanks{e-mail: jboley2@niu.edu}\\ %
        \scriptsize Northern Illinois University %
\and Maoyuan Sun\thanks{e-mail: smaoyuan@niu.edu}\\ %
     \scriptsize Northern Illinois University %
}

\abstract{
User trust is a crucial consideration in designing robust visual analytics systems that can guide users to reasonably sound conclusions despite inevitable biases and other uncertainties introduced by the human, the machine, and the data sources which paint the canvas upon which knowledge emerges.
A multitude of factors emerge upon studied consideration which introduce considerable complexity and exacerbate our understanding of how trust relationships evolve in visual analytics systems, much as they do in intelligent sociotechnical systems.
A visual analytics system, however, does not by its nature provoke exactly the same phenomena as its simpler cousins, nor are the phenomena necessarily of the same exact kind.
Regardless, both application domains present the same root causes from which the need for trustworthiness arises: Uncertainty and the assumption of risk.
In addition, visual analytics systems, even more than the intelligent systems which (traditionally) tend to be closed to direct human input and direction during processing, are influenced by a multitude of cognitive biases that further exacerbate an accounting of the uncertainties that may afflict the user's confidence, and ultimately trust in the system.

In this article we argue that accounting for the propagation of uncertainty from data sources all the way through extraction of information and hypothesis testing is necessary to understand how user trust in a visual analytics system evolves over its lifecycle, and that the analyst's selection of visualization parameters affords us a simple means to capture the interactions between uncertainty and cognitive bias as a function of the attributes of the search tasks the analyst executes while evaluating explanations.
We sample a broad cross-section of the literature from visual analytics, human cognitive theory, and uncertainty, and attempt to synthesize a useful perspective.
} 

\begin{document}

\maketitle

\section{Introduction}

In the age of ``big data'', trustworthiness in computing has become a complex, multi-faceted challenge for system designers and developers.
The rise and advancement of intelligent sociotechnical systems in particular has compounded the trust issues facing responsible, ethical software vendors; as modernized high-traffic and high-demand computational ecosystems increasingly rely on intelligent agents to aid human decision makers.
The most visible and controversial of these systems, such as those used in recidivism and credit scoring \cite{biddle2020predicting,mirtalaie2012trust}, often have a high degree of impact and carry significant implications and consequences for society \cite{wang2019designing,bellotti2001intelligibility}.
The issue of user trust in the data, models and processing of increasingly large datasets---particularly when using opaque models such as deep neural networks---has its origins within the context of automation, where the human is generally excluded or only minimally ``in the loop''.
But as the sophistication of intelligent systems and the scope of the problems they solve have evolved, so too has the need to bring humans into decision-making processes originally belonging entirely to the automated system.
At the same time, the HCI community has largely concerned itself with enabling more effective communication between human and machine, introducing more sophisticated interactive methods and ideas which have brought about several frameworks describing human-AI collaboration \cite{cai2019human}.

The themes of accountability and trustworthiness have resonated quite clearly within the visual analytics community, where both the human and the machine intersect to cooperatively steer analysis in increasingly collaborative settings (e.g., \cite{liu2021mtv}).
At the high level, the user or analyst guides the process by requesting transformations of data into useful information presented as visualizations, from which hypotheses can be formulated and tested.
At the low level, the machine responds to the user, performing the transformations of raw or preprocessed data, and also the mapping of processed results to a graphical syntax and grammar that it can use to construct meaningful visualizations.
The software industry is rife with examples of systems that fail because they are unreliable or otherwise unable to meet the needs of their users.
Collaborative, AI-backed systems are as susceptible to poor design and implementation decisions as any, if not more so because of their complexity.
But in the visual analytics domain there is even more at stake.
Responsible developers do not \textit{only} need to know when their systems are failing the users;
the analyst's trust in the tools they are expected to use is an absolutely crucial component of knowledge generation, and any visual analytics tool ignored and excluded from knowledge generation has little value.
Worse, they can be harmful if the analyst actually trusts them when they should not or cannot avoid using them \cite{sacha2015role}.

Circumventing the limitations of human cognition is at the very heart of both visual analytics and the use of visualizations in general.
Hilbert et al. supply an extremely informative discussion in \cite{hilbert2012toward} concerning noisy mechanisms in human cognitive processing, which we will later outline.
Prior to their work, over half a century of scholarship had already revealed an impressive catalog of both biases and heuristic shortcuts in human cognition, which Tversky and Tahneman compile and discuss at length in \cite{tversky1974judgment}.
Those works collectively describe and illustrate the built-in heuristics used in low-level cognition during human reasoning tasks, all designed to work around cognitive limitations. For example, Simon's ``bounded rationality'' \cite{simon1955behavioral,simon1956rational} was an early attempt to describe low-level cognitive shortcuts the human brain uses to overcome limited short-term memory and imperfect recall.
Several works have since enhanced the discussion to include computational aspects such as limited short-term storage capacity and epistemic "noise" introduced by imperfect long-term storage and recall mechanisms \cite{goldstein2002models,kahneman1982simulation,shah2008heuristics}.

We can therefore expect that such built-in cognitive mechanisms will lead to specific patterns or tendencies in how visual information is processed with some degree of consistency across users.
Hilbert et al. in \cite{hilbert2012toward} compellingly argue that \textit{all} information processing in human reasoning---from extraction to interpretation, and then clear on from reasoning with abstractions to the formulation of conclusions---is inherently noisy as an inevitable consequence of what is ultimately an ecologically competitive, but not perfect, evolutionary design.
It is reasonable to suppose that biased information processing likewise exerts biases on hypothesis formation and knowledge generation.
If low-level cognitive bias generalizes across users then so too may the ways in which a user's conclusions are influenced as low-level, attentional bias maps to high-level reasoning bias in what, to use a computation-inspired metaphor, we could consider a complete human cognitive pipeline for information discovery and symbolic reasoning.


Within the machine learning domain, Bhatt et al. \cite{bhatt2021uncertainty} contend that the measurement and communication of uncertainty is essential in building models that a user can usefully incorporate into their decision-making processes.
At first glance there may seem to be some barriers to translating their conclusions directly to the visual analytics space.
First, their discussion is constrained to uncertainty of outcomes as singletons---that is, one-time transformations of inputs to outputs---as a result of machine learning processing, instead of a sequence of interactions where the machine responds directly to user feedback to intermediate results.
Second, users of visual analytics typically have considerably more freedom to select the parameters of the analyses (e.g., type of processing or algorithm, type of graph, etc.) through which they can explore a data space.
Finally, there is typically a mismatch between the types of tasks being performed; visual analytics is \textit{explorative} while traditional machine learning algorithms simply \textit{make decisions}; for example, deduce a class or make a prediction based on features in the data learned solely through training algorithms.

However, over the past several years a similar concern for the effects of uncertainty in data and processes has emerged in the visual analytics and visualization communities, arriving at similar conclusions \cite{correa2009framework,sacha2015role}.
It is additionally worth noting that while Bhatt et al.'s work is rooted in the explainable AI (or XAI) context, visual analytics and machine learning have become increasingly intertwined over the past several years in general \cite{endert2017state,liu2017towards,yuan2021survey,sacha2018vis4ml,spinner2019explainer}.
Algorithmic, transparency and explainability certainly have their place in minimizing the user's uncertainty in the transformations performed during the analytic process, \textit{particularly} where machine intelligence is woven into the fabric of the visual analytics process, be it as a simple data source or analysis tool provider.

In this paper we propose that trust in the knowledge generation process is influenced by uncertainty that propagates from data into knowledge discovery.
Noise and errors from biased data and potentially biased processing are incorporated on the machine side into the visualizations the analysts explore to discover gain new insights.
We also take the position that cognitive bias in both foraging and the formation of explanatory associations will tend to degrade the usefulness of the analytic model---the collective interactive processes and strategies executed by the analyst to generate insight \cite{sacha2015role}.
In regards to trust calibration, we additionally maintain that the analyst's overall perception of the utility of their models within the context of the problems they solve greatly influences when and to what degree a visual analytics system is engaged, and that a lack of useful or correct results harms user trust.

We begin by presenting a survey of the literature that covers a broad range of relevant topics in sections 2, 3, and 4. In section 5 we present research specific to models of human cognition through which we build our case for the exploration of how uncertainty propagates from visualizations on the machine side into human reasoning about the data being explored, through the mechanisms of cognitive bias. We briefly discuss the implications and research questions introduced by our work in section 6 and conclude in section 7.
\section{Trust Calibration in the Knowledge Generation Process}

Definitions of trust vary widely within the fields of computer science, sociology, philosophy, and beyond. 
While traditionally such discussions center on trust between or among humans, within the past few decades, human-AI trust has come into the spotlight not only for researchers, but also for professional and casual audiences interacting with AI in their work or leisure activities. 
Before we can examine what trust means in the context of visual analytics (to say nothing of AI-driven visual analytics), we must consider different conceptualizations of human-AI trust as a whole.

In a survey of research concerning trust in AI \cite{vereschak2021evaluate}, Vereschak et al. highlight three aspects central to many definitions of human-AI trust throughout the literature. 
The first is a conceptualization of trust as an \textit{attitude} rather than a behavior. The second is a sense of \textit{vulnerability} as the user accepts that the AI might not yield a favorable result. 
The third is the presence of \textit{positive expectations}; the user believes that the AI will produce a beneficial outcome.

Vereschak et al. \cite{vereschak2021evaluate} additionally identify two main perspectives on the distinction between trust and distrust in the literature. 
From one viewpoint, trust can be conceptualized as a single continuum from trust to distrust. 
A recent version of such a continuum can be found in \cite{han2020beyond}, where distrust is placed on the spectrum adjacent to ``undistrust'' (that is, suspicion, or a lower level of distrust with an overall negative expectation for the AI's behavior), followed by the more positive notions of ``untrust'' and trust, respectively. 
From another standpoint, trust and distrust exist on separate axes; high levels of trust do not imply low levels of distrust, and vice versa. 
This perspective suggests a more granular view of trust as a multidimensional phenomenon; a user may have faith in one aspect of an AI system, for example, but not another. 
Jacovi et al.'s recent formalization of (multi-)contractual trust extends these concepts into a space defined by user objectives \cite{jacovi2021formalizing}; each \textit{contract}, or ``useful'' AI functionality, carries its own trust-distrust axis.

Trust may be thought to calibrate on a continuum by the degree to which outcomes conform to user expectations, as the user engages with the system over time.
Positive contributions to the perception of trustworthiness calibrate trust towards the positive end of the spectrum, and vice versa negative perceptions quickly move calibration in the opposite direction.
We may further conceptualize zones on the ends of the spectrum beyond which trust is more difficult to calibrate towards the opposing end \cite{han2020beyond, marsh2009examining}. A \textit{threshold of cooperation} and \textit{limit of forgivability} demarcate the boundaries of these zones, in which the user trusts or distrusts the AI system to the point where the user is either willing to engage with the system cooperatively or unwilling to believe that the system is working towards their objectives, respectively.

Several works have tackled the processes at work in knowledge generation, in both the general sense and specific application domains \cite{sacha2014knowledge,antonelli2018external,pohl2012user,sacha2015role}.
A general model of knowledge generation within the visual analytics domain was developed in \cite{sacha2014knowledge}, proposing a complex interplay between processing on the machine side and knowledge generation on the human side.
The essential components of this model are the loops through which the analyst gathers information and tests hypotheses:

\begin{enumerate}
    \item \textbf{Exploration loop}, where the analyst interacts with the system to generate observable outcomes to elicit insights leading to the formation of hypotheses.
    \item \textbf{Verification loop}, where the analyst confirms or rejects hypotheses, steering the direction of the exploration loop.
    \item \textbf{Knowledge generation loop}, where the analyst generates new knowledge through information foraging and hypothesis testing.
\end{enumerate}

Outside of trivial tasks such as initialization and interface configuration, these loops capture the majority of the analyst's interaction with the system, during which a two-way flow information is conducted.
Beyond ``one-off'' initial calibrations of trust, such as from favorable impressions of the developer or interface design, calibration of the analyst's trust predominantly occurs during the execution of activities within these loops.
Indeed, according to Sacha et al. in \cite{sacha2015role}, trust calibration occurs primarily during the knowledge generation process (i.e., the knowledge generation loop), along several dimensions originally discussed by Muir in \cite{muir1987trust}.

Muir additionally proposes three essential components situated outside the knowledge generation process that heavily influence trust-building:
1) Expert knowledge (i.e., domain expertise),
2) Technical facility, and
3) Routine performance (i.e., development of rote proficiency with a system).
As evidenced by accounts of expert problem-solving in e.g. \cite{klein2007data}, there is a significant difference between expert and novice problem-solving knowledge; the expert has background and contextual information outside of the pipeline that helps them steer their analysis around uncertainties that hinder the less-experienced analyst.
Expertise with a system, or visual analytics systems in general, also affects search behaviors though mostly in the form of stream-lining of the overall analysis.
\section{Uncertainty in Visual Analytics}

The study of uncertainty in both computation and human cognition has a long and varied history.
To date, the majority of the work has focused on the computational aspects.

\subsection{Sources of Uncertainty}

From \cite{zuk2007visualization} (as cited in \cite{sacha2015role}), uncertainty may be understood as a "composition" of several different concepts:

\begin{itemize}
    \item \textbf{Error.} Deviation of data points from their true values (e.g., outliers).
    \item \textbf{Imprecision.} Lower resolution of datapoints relative to that needed for the analysis task.
    \item \textbf{Accuracy.} Tightness of the ranges within which values tend to fall relative to true values.
    \item \textbf{Lineage.} Origin and evolution of the data (i.e., provenance).
    \item \textbf{Subjectivity.} Qualitative judgement/interpretation injected into the data.
    \item \textbf{Non-specificity.} Vagueness in the values assigned to an attribute of a data point. We alternately propose that, qualitatively, the concept includes the degree to which attributes of data points fail to meaningfully distinguish one from another as related to the performance of a task (i.e., unsuitability of selected features).
    \item \textbf{Noise.} Background clutter created by random processes that are either not explainable or not relevant to the task
\end{itemize}

Based on Sacha et al.'s discussion in \cite{sacha2015role}, we infer that imprecision, subjectivity and non-specificity especially confound user mental models, as a deficit of definitive incorporation of information into hypothesis testing. This is further complicated by the fact that the uncertainty in data sources has probabilistic characteristics, typically falling within distributions \cite{hullman2018pursuit,kale2020visual} that are not readily apparent and at worst, may give the appearance of concealing deeper information when in fact laying false trails.

Frederico et al. in \cite{federico2017role} give a more exacting account of where in the knowledge  generation process uncertainty is most likely to manifest or become injected, including 1) Automatic analysis (machine-side processes), visualization, and the interaction methods. Automatic analysis is not only subject to potential algorithmic bias, but in general statistical errors carried through from the raw datapoints is amplified. The communication of information can also confound uncertainty when visualizations are inappropriately used or poorly constructed (as discussed at length by Lo et al. in \cite{lo2022misinformed}) distracting from or obscuring important/salient information. Finally, interaction methods introduce uncertainty when they misalign with potential ``paths'' of investigation that would efficiently lead to useful new information, i.e., complicate the means of investigation (e.g., added steps), divert the user from identifying useful patterns, or otherwise hinder the analyst's mental processing.

Looking to such works as \cite{lo2022misinformed}, we may also find consideration of the data sources, where both raw and preprocessed data are recognized to introduce bias, as well as other types of statistical error. More subtly, views of data , or filters applied to raw or derived data that e.g. conform data to a model (i.e., dropping irrelevant features), or filter samples where one or more features do not fall into specific ranges of values, may potentially obscure relevant patterns \cite{huang2018big}.

\subsection{Propagation of Uncertainty}

Accounting for uncertainty is by far simplest in the analytic system:
Uncertainty enters via data sources and is amplified via processing that first incorporates data into a model and then interprets models into a visualization \cite{sacha2015role}.
A pipeline describing the capture of uncertainty in visualization is presented by Griethe and Schumann in \cite{griethe2006visualization}, however this pipeline also captures subsequent entry points of uncertainty during the filtering, mapping and rendering stages.
As implied in \cite{griethe2006visualization}, each is influenced by a set of parameters that are required for meaningful, task-relevant interpretation of the data, collectively encoding a set of assumptions and/or knowledge artifacts.
Derived from prior propositions, expertise and information, all of which carry over some uncertainty introduced during prior processing.
Assumptions tend to rely on heuristic generalizations of human knowledge constructs, making them subject to bias.

Uncertainty is somewhat more difficult to account for on the human side of the process.
Obviously, there is no direct or practical way to measure the analyst's uncertainty on a step-by-step basis.
However, it is possible to identify where uncertainty manifests.
Zuk and Carpendale's work in \cite{zuk2007visualization} presents knowledge constructs in the context of uncertainty visualization, citing three primary types of constructs as discussed in \cite{cook2005illuminating}:

\begin{enumerate}
    \item \textbf{Arguments.} In the words of Thomas and Cook (in \cite{cook2005illuminating}), the "logical inferences linking evidence and other reasoning artifacts into defensible judgments of greater knowledge value". As set forward in \cite{zuk2007visualization}, uncertainties contribute to ill-defined and poorly structured problems, creating difficulties with devising useful or valid argumentation by which to solve them as problem specifications may be unclear relevance of data and representations become nebulous, among other problems (the reader may refer to \cite{bruggen2003cognitive} for a deeper discussion).
    \item \textbf{Causality.} The cause-effect relationship between events, where actions contribute to resulting states and responsive actions. Perception of causality is highly temporalized such that co-occurrent events may be misconstrued as causal. Building on \cite{zuk2007visualization}, we additionally note that causality may be falsely implied by temporal characteristics of a visualization chain, creating false leads and maladaptively distorting the analytic model. User preconceptions may additionally bias towards assumptions of causality even when diagnostic information establishes otherwise [15].
    \item \textbf{Models of Estimation.} Allow for the mapping of concrete data to useful abstractions. Uncertainties may be introduced during.
\end{enumerate}
\subsection{Effects of Uncertainty on Trust Calibration}

Sacha et al. posit in \cite{sacha2015role} that trust calibration is influenced by the analyst' \textit{awareness} of uncertainty, indicating that recognized uncertainty tends to correctly calibrate trust, while that remaining undetected will generally miscalibrate trust.
However, mistaken uncertainty, whether as a result of uncertainty bias or miscommunication is also possible, and will tend to miscalibrate trust.
Miscalibrated trust ultimately leads to greater bias and error in the results, detrimentally impacting the analyst's trust in the entire analytic process over time.
Miscalibration disrupts the knowledge generation process in several ways, not least among them the distrust for the system as analyses tend to generate results that are incorrect or have low utility, and possibly even degradation of the user's trust in their own competence in the worst case, depending on their domain expertise.

According to Sacha et al. \cite{sacha2015role}, user trust will tend to be high in two cases:
1) When there are little to no uncertainties and the analyst is aware, in which case trust is properly calibrated; and
2) When there are uncertainties and the analyst is unaware, in which case trust is micalibrated.
The analyst may also mistakenly perceive uncertainties when there are none, in which case trust is lower than it should be and therefore miscalibrated.
If there are uncertainties but the analyst is \textit{mistaken} about what they are, whether as a result of bias or miscommunication, trust will likewise become miscalibrated and overall trust will be lower should the problems persist.
It is important to note, however, that trust may not \textit{necessarily} be miscalibrated in the short term, as the overall level of calibration might be appropriate.
Unfortunately, the nature of the bias and errors introduced may surprise the user, reducing overall trust in the system.

\section{Cognitive Bias in Visual Analytics}
\label{sec:cog-bias}

Despite a common negative connotation, cognitive bias has the adaptive quality of enabling rapid decision-making in the absence of detailed information or when time constraints preclude rigorous analysis \cite{wall2017warning,dimara2018task}.
However, in the analytic space, cognitive bias often works against the analyst when superficial patterns in the data obscure those that are deeper and more meaningful.
The effects of cognitive biases are both fairly consistent and predictable, and manifest as traceable interactive behaviors \cite{wall2017warning}.
Within the context of intelligence analysis, Heur in \cite{heuer1999psychology} identifies several types of cognitive bias that emerge during evaluation:

\begin{enumerate}
    \item \textbf{Vividness criterion.} The analyst attends to data and information that is more salient (i.e., specific, personally relevant) than that which is vague or nebulous.
    \item \textbf{Absence of evidence.} Evidence that is missing tends to be overlooked in favor of evidence that is present.
    \item \textbf{Oversensitivity to consistency.} Analysts build or otherwise favor hypotheses that are (superficially) supported by the most data.
    \item \textbf{Coping with evidence of uncertain accuracy.} Analysts tend to overlook partial accuracy of a hypothesis, wholly accepting or rejecting it instead.
    \item \textbf{Persistence of impressions based on discredited evidence.} Analysts tend to carry over conclusions reached prior to the discrediting of hypotheses, regardless of its validity.
\end{enumerate}

Dimara et al. extend the discussion of cognitive bias in \cite{dimara2018task}, introducing a taxonomy of documented bias grouped by task type. They discuss an \textit{estimation task}, in which users engage in the assessment of some value relative to a question or problem-solving step, as well as a \textit{decision task}, which consists of the selection of an option among many. Also discussed is a \textit{hypothesis assessment task}, in which the user evaluates the hypotheses formulated during the knowledge generation loop. Finally, a \textit{causal attribution task} is introduced, where users attempt to infer a cause-effect relationship among events or pieces of evidence.

Dimara et al. further observe that types of cognitive bias coalesce around a set of fundamental characteristics (or ``flavors''), regardless of the type of task: Association, baseline, inertia, outcome, and self-perspective. Associations are the perceived connections between evidence which naturally predispose the user towards assumptions incorporating those connections. Assumed baselines (with or without supporting evidence) predispose the user towards a specific standard for comparison that influences further analysis. In the group of biases grouped under outcome, the central theme is expectation creating cognitive influence that steers the analysis. In the self-perspective category, analysis in influenced (or constrained) by the particular viewpoint of the analyst, whether as a result of domain expertise or uninformed assumptions.

From the preceding discussion it also emerges, though indirectly, that prior patterns of interaction (i.e., habits) and reinforced perceptions of the relevance of subsets of data points or attributes may also contribute to observable higher-level forms of bias; in other words, biases can be learned.
These manifest as patterns in what types of data are attended to most frequently or what types of analytic tasks are most frequently undertaken.

\section{The Interplay between Uncertainty and Cognitive Bias in Sensemaking}

Based on the conclusions offered by Sacha et al. in \cite{sacha2015role}, we expect that cognitive biases may misdirect the user into information seeking behaviors that may not be consistently effective or useful.
Those behaviors may either create incorrect perceptions of uncertainty as the analytic model degrades or deflect the analyst's attention from uncertainties that are important to the proper calibration of trust.
This directly implies that cognitive bias may have detrimental effect on the analyst's overall trust in the system or knowledge generation process.

Klein et al.'s view of sensemaking as a data-framing problem from \cite{klein2007data} potentially offers a means to describe user analytical reasoning in the context of a network of relationships between propositions and premises.
Klein et al. describe these networks as ''frames'', where information extracted from visualizations is used to formulate propositions and gather supporting (or contradictory) evidence that strengthen or weaken inferred relationships.
Klein et al. do not provide a clear definition of the term ``data''; we therefore adapt their term to describe any discrete piece of information derived from actual data (e.g., a trend in a line graph, clustering in a scatterplot, etc.) that the user incorporates into their reasoning.

Similar to Correa et al.'s conception of uncertainty in data sources from \cite{correa2009framework}, we propose that it is possible to capture uncertainty within a user's frame through Gaussian mixture models.
This would be convenient, as it would allow us to think of ``Kleinian'' frames as a type of Gaussian network that changes over time, presumably with a central set of propositions forming more stable regions of the network that would tend to converge relatively early on.
Restructuring would occur as new relationships are inferred, old relationships are discarded, and new propositions are discovered.

Klein additionally reviewed 120 case studies in \cite{klein2013seeing}, noting that the "anchors", or central propositions of an explanatory framework shift as new insights are generated.
These anchors directly steer the analysis and guide the analyst's attention in the subsequent data and information foraging activities, until such a time as contrary or unaccounted-for evidence spurs further investigation, potentially leading to a shift or restructuring in the framework.
We additionally observe that under ideal conditions, the frequency and degree of shifts over time should stabilize as the analyst converges upon a solution.

It therefore seems reasonable to propose that 1) the trend of such convergence reveals much about the degree of uncertainty in the analyst's reasoning, and 2) the characteristics of the information foraging activities (e.g., number of steps, number of unrelated data sources, the amount of time spent using a visualization before it is discarded, etc.) reflect the analyst's uncertainty regarding a proposition under investigation.
We can think of this in terms of the parameters of the search: If the analyst reuses a significant portion of the parameters used in previous searches regarding the subset of entities of the frame targeted by investigation, then investigation is likely confirmatory.
However, if relatively few are employed, then it is likely that the investigation either seeks to resolve conflicting new information, is following a loosely-related parallel, supportive line of reasoning, or the analyst is pursuing a new conception (i.e., expanding upon the frame) that happens to include some common parameters.
For our purposes, we are most interested in the first case.
We therefore propose that a reasoning process can be approximately described as a series of parameter selection problems, and that the correlations between the frequencies with which parameters are selected across problem instances reveal useful information in regards to the degree of uncertainty in the analyst's reasoning.

\section{So Where Does This Leave Us?}

The problem of tracing uncertainties in visualization through the analysis process is extremely difficult, especially given that cognitive bias is not likely to present in exactly the same way under all conditions by all users.
The authors are unaware of any significant, direct undertaking to rigorously investigate how cognitive biases affect the uncertainties that propagate from the machine into the analyst's decision making.

However, we can make some educated guesses.
Based on the discussion in \cite{sacha2015role}, we should expect that cognitive biases likely misdirect the user into information seeking behaviors that are only partially effective at best.
We can suppose those behaviors could also create incorrect perceptions of uncertainty, assuming the system attempts to communicate them (e.g., \cite{correa2009framework,zuk2007visualization,hullman2018pursuit}, etc.).
In any case, it is likely to deflect the analyst's attention from important information, possibly calibrating user trust downwards.
If the analyst remains unaware of it then their trust will likely not be appropriately calibrated, with potentially serious consequences. 

While it is possible to model uncertainty as it propagates and aggregates during analysis on the machine side, we lack a unified model of the propagation of uncertainty through analytic reasoning.
As noted in \cite{wall2017warning}, the purpose of the visual analytic process is to reduce the uncertainty introduced by the kinds of reasoning tasks discussed in \cite{hilbert2012toward}.
While visual analytics has proven to handily mitigate many of the errors introduced through human reasoning by externalizing both memory channels and significant transformations of information, human cognitive bias is still a significant factor \cite{wall2017warning}. It is our belief that a better understanding of how error transmitted to the human \textit{through} the visualizations influences information gathering behaviors, is required.

A related question is how the influence of such biases could be measured.
Our proposal ultimately assumes that observable characteristics of the interactions reveal useful information about the analyst's internal mapping of information to evidence for or against a hypothesis, and that there is a definite correlation between the analyst's overall trust over time and their specific foraging behaviors.
Wall et al.'s work in \cite{wall2017warning} does handily establish that cognitive biases can be detected through interaction traces, but how would the data recorded in these traces correlate with the analyst's foraging strategies, and what can be inferred about user trust from them?
These and similar questions must be assessed before we can even begin to determine whether this is a line of inquiry worth pursuing. 
\section{Conclusion}

In this article we have argued the need for an accounting of how uncertainty propagates from data sources through to the extraction of information and hypothesis testing as a requirement to a deeper understanding of how user trust in a visual analytics system evolves over its lifecycle.
We presented a survey of a segment of the literature regarding uncertainty and cognitive bias in visual analytics, focused on those works which seem to collectively offer the best clues towards the creation of a useful theoretical framework for investigating the propagation of mechanical uncertainty through cognitive bias into resolution of the analyst's uncertainty in their reasoning.
Our work continues as we gather more information from the literature and refine and extend our models.


\acknowledgments{
This research is supported in part by NSF Grant IIS-2002082 and the Research and Artistry Opportunity Grant from Northern Illinois University.}

\bibliographystyle{abbrv-doi}

\bibliography{refs}
\end{document}